\documentclass{aa}  

\usepackage{graphicx}
\usepackage{txfonts}
\usepackage{lipsum}
\usepackage{subcaption}         
\usepackage{lscape}             
\usepackage{placeins}           
\usepackage{orcidlink}     

\newcommand{\psrcomplete}{PSR J2339-0533}
\newcommand{\psr}{J2339}

\newcommand{\fluxcgs}{erg\,s$^{-1}$\,cm$^{-2}$}

\graphicspath{{./}{figures/}}

\begin{document}

   \title{Evidence for optical pulsations from a redback millisecond pulsar}



   \author{A.~Papitto\inst{1} \orcidlink{0000-0003-2810-2394}
        \and F.~Ambrosino\inst{1}  \orcidlink{0000-0001-7915-996X} \and  M.~Burgay\inst{2} \orcidlink{0000-0002-8265-4344} \and R.~La Placa\inst{1} \orcidlink{0000-0003-2810-2394} \and C.~J.~Clark \inst{3,4} \orcidlink{0000-0003-4355-3572}\and C.~Ballocco \inst{1,5} \and G.~Illiano \inst{6} \orcidlink{0000-0003-4795-7072} \and C.~Malacaria\inst{1} \orcidlink{0000-0002-0380-0041} \and A.~Miraval Zanon\inst{7} \orcidlink{0000-0002-0943-4484} \and A.~Possenti\inst{2} \orcidlink{0000-0001-5902-3731} \and L.~Stella\inst{1} \orcidlink{0000-0002-0018-1687} \and A.~Ghedina\inst{8} \orcidlink{0000-0003-4702-5152} \and M~Cecconi\inst{8}\and F.~Leone\inst{9} \orcidlink{0000-0001-7626-3788} \and M.~Gonzalez\inst{8} \and H.~Perez Ventura\inst{8} \and M.~Hernandez Diaz\inst{8} \and J.~San Juan\inst{8} \and H.~Stoev\inst{8}}

   \institute{INAF Osservatorio Astronomico di Roma, via di Frascati 33, I--00078, Monteporzio Catone, Roma, Italy
            \and {INAF – Osservatorio Astronomico di Cagliari, Via della Scienza 5, I-09047 Selargius (CA), Italy}
            \and {Max Planck Institute for Gravitational Physics (Albert Einstein Institute), D-30167 Hannover, Germany}
            \and {Leibniz Universit{\"a}t Hannover, D-30167 Hannover, Germany}
            \and {Dipartimento di Fisica, Sapienza Universit\`a di Roma, Piazzale Aldo Moro 5, I-00185 Rome, Italy}
            \and {INAF–Osservatorio Astronomico di Brera, Via Bianchi 46, I-23807, Merate (LC), Italy}
            \and {ASI - Agenzia Spaziale Italiana, Via del Politecnico snc, 00133 Roma, Italy}
            \and {Fundaci\'on Galileo Galilei - INAF, Rambla Jos\`e Ana Fernández P\`erez, 7, E--38712 Bre\~na Baja, TF - Spain}
            \and {Universit\`a degli Studi di Catania, Via S. Sofia, 64, I--95123 Catania, Italy}}
   \date{Received \today }
 \abstract{Recent detections of optical pulsations from both a transitional and an accreting millisecond pulsar have revealed unexpectedly bright signals, suggesting that the presence of an accretion disk enhances the efficiency of optical emission, possibly via synchrotron radiation from accelerated particles. In this work, we present optical observations of the redback millisecond pulsar PSR J2339–0533, obtained with the SiFAP2 photometer mounted on the Telescopio Nazionale Galileo. Data accumulated  during the campaign with the longest exposure time ($12$~hr) 
 suggest that its $\sim$18 mag optical counterpart exhibits pulsations at the neutron star’s spin frequency. This candidate signal was identified by folding the optical time series using the pulsar ephemeris derived from nearly simultaneous observations with the 64-m Murriyang (Parkes) radio telescope. The detection significance of the candidate optical signal identified in those data lies between 2.9 and 3.5 $\sigma$, depending on the statistical test employed.  The pulsed signal has a duty cycle of $\approx 1/32$, and the de-reddened pulsed magnitude in the V band is $(26.0 \pm 0.6)$ mag. At a distance of 1.7 kpc, this corresponds to a conversion efficiency of $\sim 3 \times 10^{-6}$ of the pulsar’s spin-down power into pulsed optical luminosity—comparable to values observed in young, isolated pulsars like the Crab, but $50$–$100$ times lower than in disk-accreting millisecond pulsars. If confirmed, these findings suggest that optical pulsations arise independently of an accretion disk and support the notion that such disks boost the optical emission efficiency. }


   \maketitle

\section{Introduction}

Millisecond pulsars (MSPs) are thought to form through a prolonged (0.1–1 Gyr) accretion phase in which a neutron star gains mass and angular momentum from a low-mass companion \citep{Alpar1982}. During this low-mass X-ray binary phase, X-ray pulsations may occur as the magnetic field funnels in-falling material towards poles \citep{Wijnands1998}. Once accretion ceases, the system transitions into a rotation-powered state, with the neutron star emitting primarily in radio and gamma rays \citep{Backer1982,Abdo2009}. A rare class of systems -- transitional millisecond pulsars -- has been observed to switch between accretion- and rotation-powered states on timescales of days to weeks, due to variations in the accretion rate (\citealt{Archibald2009, Papitto2013}).
Recently, optical pulsations at the spin period of the neutron star have been detected from both a transitional and an accreting MSP using the fast optical photometer SiFAP2 on the 3.6-m Telescopio Nazionale Galileo (TNG; \citealt{Ambrosino2017,Ambrosino2021}; see also \citealt{Zampieri2019} and \citealt{Karpov2019}). The unexpectedly high brightness of these optical pulsations challenges traditional models based solely on rotation- or accretion-powered mechanisms, suggesting that the interaction of the pulsar’s relativistic wind with the accretion disk plays a key role \citep{Papitto2019, Veledina2019, Illiano2023}. Rotation-powered MSPs in compact binaries share similar spin and orbital characteristics with transitional MSPs and are ideal control cases to test the origin of optical pulsations. So far, only an upper limit was reported  \citep[$m_{\rm g}<25$~mag;][]{strader2016}.

\begin{figure*}[t!]
    \centering
     \resizebox{\hsize}{!}
    {\includegraphics {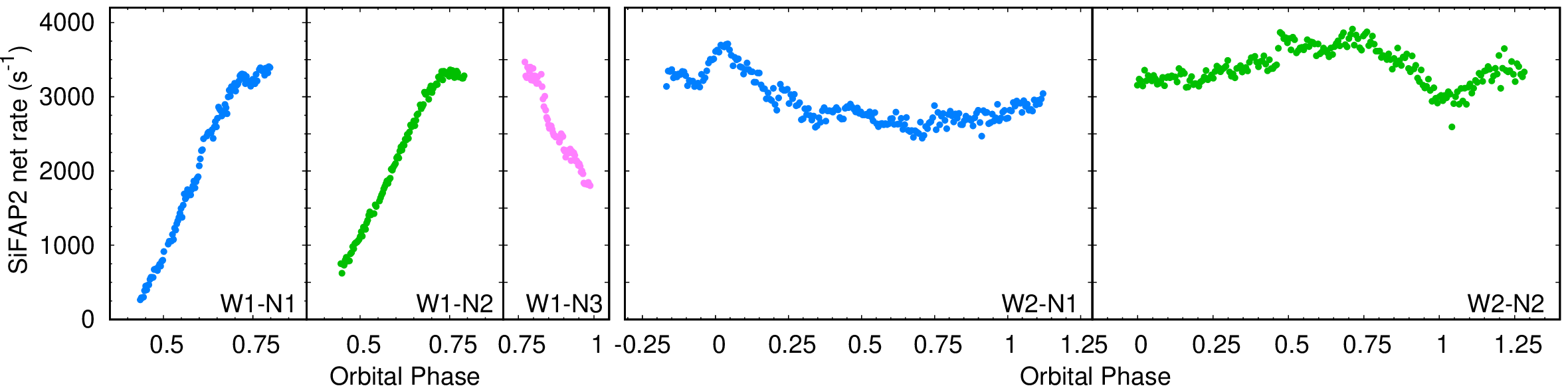}}
     \caption{SiFAP2 net light curves of {\psr} during the different nights (labelled as N$i$) of the two observing campaigns W$i$ (see text for details on the background estimation). The orbital phase was evaluated considering the timing solution listed in Table~\ref{table:psr}. 
     }
      \label{fig:fig1}
\end{figure*}

In this paper, we report on SiFAP2/TNG observations of PSR J2339–0533 (hereafter J2339). Initially identified as a candidate MSP in a $4.6$~hr orbit around a $\sim 0.3 \,M_\odot$ companion based on its gamma-ray, X-ray, and optical properties  \citep{Romani2011,Kong2012}, the detection of 2.8 ms radio and gamma-ray pulsations later confirmed its nature \citep[][and references therein]{Ray2020}. Its radio emission is visible only near inferior conjunction, likely due to obscuration by intra-binary ionised material, as often observed in redback systems \citep{Archibald2009}. The optical counterpart displays brightness variations from $m_g = 18$ to 23 across the orbit \citep{Romani2011,Yatsu2015,Kandel2020}.

\section{Observations and data analysis}

SiFAP2, a non-imaging fast photometer \citep{ghedina2018}, observed {\psr} on August
10–12, 2021 for an exposure of 16.9~ks (W1 in the following) and on
August 26-27, 2022 for 43.2~ks (W2). The observations were in  white light (320–900 nm), with response peaking at $\simeq 450$~nm.
The two SiFAP2 detectors recorded with an 8~ns resolution the times of arrival of individual photons coming from a field of view of 7$\times$7~arcsec$^2$ around the source position, and in a source-free region 4.2~arcmin away from the source, to measure the sky background. 
During W1, we also estimated the background close to the source by tilting the pointing direction of the telescope a few times by a $\sim 20$ arcsec for $\sim$100~s (so-called nodding). The count rate observed by the two detectors during these intervals differed by $\approx 3.5\times10^3$~s$^{-1}$, most likely due to spatial variations in the sky background. To produce the background-subtracted light curves shown in the left panel of Fig.~\ref{fig:fig1}, we rescaled the sky count rate to match the rate of the other detector during nodding intervals (i.e. when it was off-source).
During W1, the count rate modulation at the 4.6~hr binary period peaks when the pulsar is at the inferior conjunction ($\phi_{\rm orb}\simeq 0.75$), resembling that in previous observations of this source \citep[$m_{\rm V}\simeq18.25$~mag;][]{Romani2011,Yatsu2015,Kandel2020}.
According to the calibration curve of SiFAP2\footnote{see \url{https://www.tng.iac.es/instruments/sifap2/}}, the detector is expected to record $(3.9\pm1.1)\times10^{3}$~$s^{-1}$ from a 
star at that magnitude,
in reasonable agreement with the observed values. Nodding was not performed during W2, preventing us from accurately determining the background close to the source. We rescaled the sky count rate by the same amount used in W1, obtaining the curves shown in the right panels of Fig.~\ref{fig:fig1}. Regardless of the uncertain background subtraction, an orbital modulation similar to that of W1 was not seen. A similarly erratic behaviour has already been reported in observations of a 
black widow pulsar \citep{Romani2011}.

We searched SiFAP2 data for a coherent signal at the spin frequency of {\psr}, first correcting the photons' arrival times to the Solar System barycentre with DE405 ephemerides. To further account for the delays introduced by the pulsar binary motion
determining the pulsar ephemeris close to the observations' epoch is crucial. Similar to other MSPs in compact binary systems, the orbital period of {\psr} varies by $\Delta P_{\rm orb}/P_{\rm orb}\simeq\mbox{a few} \times 10^{-7}$ on timescales of weeks/months, most likely due to companion gravitational mass quadrupole moment changes \citep{Pletsch2015}. Our SiFAP2 observations 
span too few orbital cycles to be sensitive to such variations.
However, unmodeled variations in the orbital period cause the epoch of passage of the pulsar at the ascending node of the orbit $T_\mathrm{asc}$ to differ by up to tens of seconds from the value predicted by using a constant orbital period solution determined years earlier. Such variations in $T_\mathrm{asc}$ can be larger than what a few orbital period-long observations are sensitive to ($\sigma_{T_\mathrm{asc}}\simeq 2.5$~s, see, e.g., \citealt{Caliandro2012}), and deserve a careful treatment.

\begin{figure}[t!]
   \centering
   \includegraphics[width=\hsize]{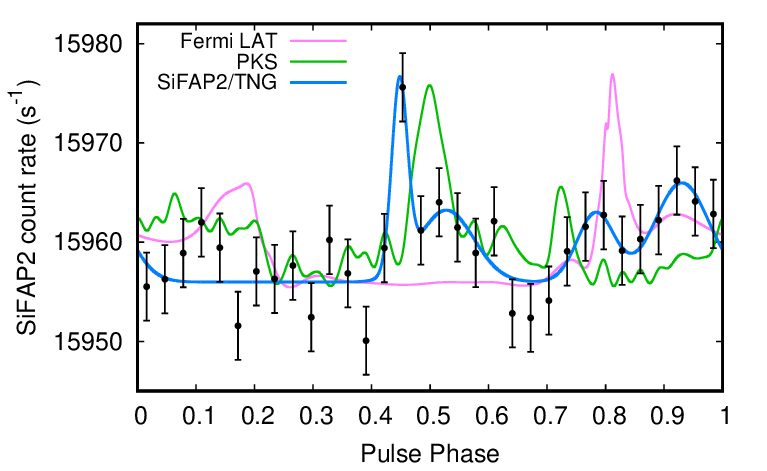}
      \caption{Phase-aligned  pulse profiles observed in W2 by SiFAP2 (black points), Parkes (green line) and Fermi LAT (magenta line; taken from 3PC catalogue, S23, and shifted assuming that the lag with the radio profile is the same as reported there).  The normalization of the radio and gamma-ray profiles is arbitrary. The reference epoch was set to align the radio peak at phase 0.5. The blue line is a fit of the SiFAP2 pulse profile with four Gaussians added to a constant level.  }
         \label{fig:fig2}
   \end{figure}

To measure the value of $T_\mathrm{asc}$ close to W2, on 2022 August 30-31, 
we observed J2339 for 10 h (two 2.5-hr observations each day) with the Ultra-Wide bandwidth Low receiver \citep[UWL;][]{Hobbs2020+UWL} of the 64-m Murriyang (Parkes) telescope. We split the  0.7--4 GHz bandwidth 
of the UWL receiver into 1-MHz-wide channels, and we measured the total intensity in search mode by recording 8-bit sampled data every 64 $\mu$s. Starting from the publicly available Fermi-LAT timing solution covering the interval MJD 54682--58826 \citep[][S23 in the following]{Smith2023}\footnote{see \url{https://fermi.gsfc.nasa.gov/ssc/data/access/lat/3rd_PSR_catalog/3PC_HTML/J2339-0533.html}}, we used the python code \texttt{SPIDER\_TWISTER} \footnote{\url{https://alex88ridolfi.altervista.org/pagine/pulsar_software_SPIDER_TWISTER.html}}, based on \texttt{PRESTO}
\footnote{\url{https://github.com/scottransom/presto}} 
\citep{PRESTO} folding routines, to find a local value of $T_\mathrm{asc}$ maximising the  signal-to-noise ratio of the folded pulsed profile. After folding the data with the newly found local $T_\mathrm{asc}$, we extracted pulse times of arrival with {\texttt{PRESTO}}'s routine {\tt{get-TOAs.py}} and refined the orbital ephemeris using {\texttt{tempo2}}
\citep{tempo2}. This allowed us to obtain the parameters listed in Table~\ref{table:psr}.
A yellow line in Fig.~\ref{fig:fig2} shows the radio pulse profile. The uncertainties in the radio timing parameters are significantly smaller than the frequency resolution afforded by our data, allowing us to perform a single trial in a signal search in SiFAP2 W2 data. 
The narrow peaks observed in the radio and gamma-ray pulse
profiles of J2339 and the low duty cycle of optical pulses of rotation powered pulsars (see, e.g., Fig. 15 in Papitto et al. 2019 for
the SiFAP profile of the Crab pulsar) motivated us to epoch-fold the SiFAP2 data using $n=32$ phase bins.
We obtained the profile plotted with black points in Fig.~\ref{fig:fig2}. 
A single bin-wide peak exceeds the average count rate by $\sim 4.5$ times its uncertainty and anticipates the radio pulse peak by $135\pm45$~$\mu$s. Although with a lower statistical significance, an excess at phases 0.8--0.9 is also seen, simultaneous to the peaks observed by Fermi LAT at gamma-ray energies.

To evaluate the statistical significance of the optical pulse, we first used the epoch folding $\chi^2$ test \citep{Leahy1983,Leahy1987}. Assuming Poissonian counting noise, the epoch folding follows a $\chi^2_{n-1}$ distribution 
when a signal is absent. However, the noise of SiFAP2 is partly correlated.
Secondary avalanches triggered by photon detection in Silicon Photomultiplier detectors
generate spurious events indistinguishable from astrophysical ones  \citep[so-called crosstalk;][]{Gallego2013,Klanner2019} and modify the distribution of noise powers. \citet{LaPlaca2025} showed that SiFAP2 noise follows a generalised Poisson distribution with a dispersion index (i.e. the ratio of variance to average) equal to $\simeq 1.15$ times what is expected from uncorrelated Poissonian noise. In appendix \ref{sec:app}, we show that a value of $r=1.16$ applies to the data considered here. To evaluate the statistical properties of the observed profile using the usual $\chi^2_{n-1}$ distribution, we then rescaled by $r$ the observed value of the epoch folding $\chi^2$, obtaining $\chi^2_{\rm W2}=64.4$. For $n=32$, the probability of observing a value equal to or higher than this in W2 in the absence of a periodic signal is $4.2\times10^{-4}$, corresponding to a 3.5 $\sigma$ confidence level.
 We evaluated the (non-background subtracted) rms amplitude of the signal as $A_{\rm rms}=\sum_j (R_j/R - 1 )^2 / n = [(\chi^2_{\rm W2}-\chi^2_{\rm noise})/N_{\gamma}]^{1/2}=(2.2\pm0.5)\times10^{-4}$. Here, $R$ and $R_j$ are the average and j-th bin count rates, respectively, $N_{\gamma}$ is the total number of photons, and $\chi^2_{\rm noise}=(n-1)$ is the $\chi^2$ expected value in the absence of a signal.  The pulse profile is marginally stronger ($\chi^2_{\rm sup}=57.2$, $A_{\rm rms}^{\rm sup}=(2.5\pm0.7)\times10^{-4}$) during the interval of orbital phases centred around the pulsar's superior conjunction ($\phi_{\rm orb}=0$--$0.5$) than during the interval centered around the inferior conjunction ($\phi_{\rm orb}=0.5$--$1$; $\chi^2_{\rm inf}=32.1$, $A_{\rm rms}^{\rm inf}<2.5\times10^{-4}$ at 3$\sigma$ confidence level). However, the shorter exposure accumulated during the latter phase interval prevented us from assessing whether the difference is significant.
 
 To employ an unbinned statistical test, 
we also applied an H-test \citep{deJager1989,deJager2010}, obtaining a value of $H^*=18.1$ for the profile observed during the whole W2. To evaluate the probability to observe a value equal to or higher than this from data containing only noise, we considered the distribution of the H values measured at 100,000 independent trial frequencies close to but not compatible with the known spin frequency (see Appendix A), obtaining a probability $p(H>H^*)=4.3\times10^{-3}$, which would correspond to a 2.9 $\sigma$ confidence level.

\begin{table}[t!]
\caption{{\psr} 64-m Murriyang (Parkes) timing parameters}                 
\label{table:psr}    
\centering                        
\begin{tabular}{l c }      
\hline\hline               
Parameter & Value \\         
\hline                      
   $\nu$ (Hz) &    346.713378637(30) \\
   $T_\mathrm{asc}$ (MJD) & 59821.68859489(36) \\
   $P_{\rm orb}$ (s) & 16683.703(8) \\
   $a_1 \sin{i} / c$ (lt-s) & 0.611668(9) \\
  
\hline                                  
\end{tabular}
\tablefoot{The signal was detected at a dispersion measure of 8.17~cm$^{-3}$~pc. The solution is referred to an epoch MJD 59817.98572011.
}
\end{table}

The absence of nodding during W2 prevented us from reliably measuring the sky contribution. In addition, the optical emission of {\psr} is largely dominated by the unpulsed contribution coming from the irradiated companion star. To estimate the pulsed flux, we then evaluated the pulsed count rate from the average rms amplitude observed during W2, $A_{\rm rms} R_{W2}=(3.5\pm0.8)$~s$^{-1}$, and used the SiFAP2 calibration curve
to obtain the corresponding magnitude $m_{\rm V}^{\rm pulse}=(26.4\pm0.6)$~mag. 
Based on the absorption column ${\rm N}_{\rm H}$ towards {\psr} estimated from HI maps \citep{HI4PI2016}, we followed \citet{Guver2009} to evaluate $0.133(5)$~mag of expected interstellar extinction in the V band. 
In addition, the average relative airmass of 1.60 during W2 gave an atmospheric extinction\footnote{\url{https://www.ing.iac.es/Astronomy/observing/manuals/ps/tech_notes/tn031.pdf}} of $0.27(1)$~mag at 450~nm. Assuming a flat spectral distribution, the pulsed magnitude corrected for extinction gives an average pulsed flux of $F_{\rm V}^{\rm pulse}=(3.1\pm1.5)\times10^{-16}$~\fluxcgs. To check the reliability of estimating the pulsed flux using the calibration curve, we measured the value obtained in the case of 2017, May 23--24 SiFAP2 high-mode observations of PSR~J1023+0038 as $(2.5^{+1.3}_{-0.9}\times10^{-14})$~\fluxcgs. Although slightly smaller, this value is broadly compatible within the uncertainties with that measured by \citet{Papitto2019}, $(4.1\pm0.2)\times10^{-14}$~\fluxcgs,  whose accuracy was granted by simultaneous spectroscopic and photometric observations to calibrate the SiFAP2 flux.

The lack of radio observations close to the epoch of W1 SiFAP2/TNG observations 
prevented us from phase-connecting the solution obtained in W2 data, and %
forced us to search for a periodic signal at the pulsar period over a $\pm 15$~s range of $T_\mathrm{asc}$ values spaced by 0.5~s, around the value predicted using the $T_\mathrm{asc}$ value listed in Table~\ref{table:psr} and the orbital period given by S23.
On the other hand, at the rate estimated by S23, the spin frequency variation from the W2 value is much lower than the frequency resolution.
The epoch folding $\chi^2$ reached a maximum of $\chi_{\rm W1, max}^{2}=52.7$ 
for $T_{\it ASC}=59437.03632$~MJD, 
not significant considering the 61 $T_\mathrm{asc}$ trials. 
The corresponding upper limit at 3$\sigma$ confidence level on the rms amplitude is $3.7\times10^{-4}$, higher than the W2 value. 
Therefore, we conclude that the lower photon statistics and the lack of a $T_\mathrm{asc}$ measurement close to W1 prevented us from reaching a sensitivity high enough to detect a signal with the same amplitude observed in W2. Taking into account the number of trials, the probability to jointly observe $\chi_{\rm W1, max}^2$ and $\chi_{\rm W2}^2$ in the absence of a signal is $1.8\times10^{-3}$.

\section{Discussion and Conclusions}

   \begin{figure}[t!]
   \centering
   \includegraphics[width=\hsize]{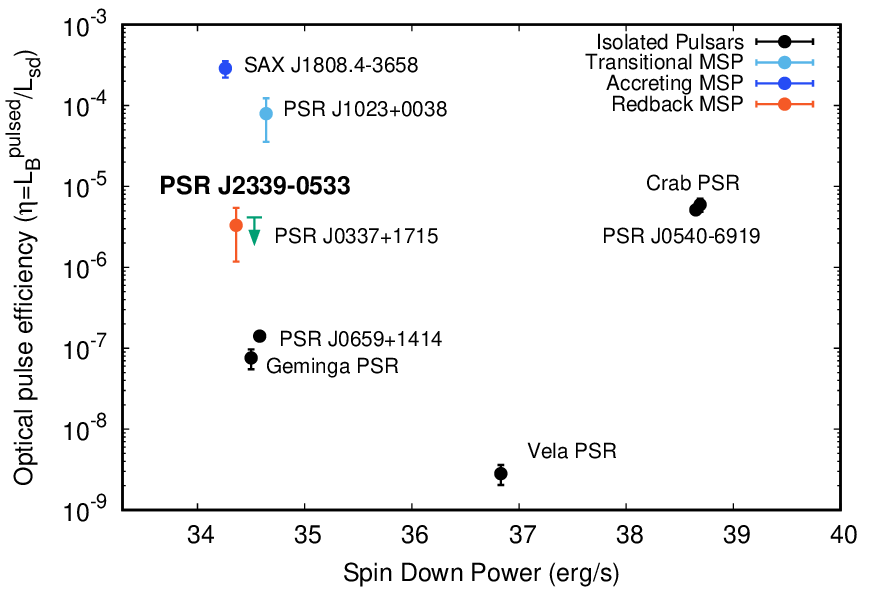}
      \caption{Ratio of the pulsed B ($\lambda_{\rm eff}=433$~nm, $W_{\rm eff}=90$~nm) luminosity to the spin down power for {\psr} (orange dot) and the sample of detected optical pulsars. Data from \citet[see Figure 3, Supplementary Table 1, and references therein]{Ambrosino2017}, \citet{Papitto2019} and \citet{Ambrosino2021}.}
         \label{fig:fig3}
   \end{figure}

Analysing SiFAP2/TNG data taken on August 26--27, 2022, we reported the detection of possible optical pulsations from {\psr} with a significance ranging between a 2.9 and a 3.5 $\sigma$ confidence level, depending on the adopted test statistics. The observed pulsed magnitude of $m_{\rm V}^{\rm pulse}=(26.4\pm0.4)$~mag is very close to the sensitivity limit that SiFAP2/TNG can achieve in a 12~hr-long exposure. 
If confirmed, the detection of optical pulsations from {\psr} would be the first from a rotation-powered MSP. 
The signal could be observed only after correction of the photon arrival times with the pulsar ephemeris; we argue below that it likely originates close to the neutron star. The signal power decreases below detectability when the epoch of passage at the ascending node is varied by $\Delta T_\mathrm{asc}\simeq 1$~s from the value measured during radio observations. This translates into an arc-length of $2\pi a_{\rm 1} (\Delta T_\mathrm{asc} / P_{\rm orb})\approx 27$~km (for $i=57^{\circ}$; \citealt{Romani2011}) around the pulsar position along its orbit. Alternatively, the signal could be produced by scattering of companion photons by a highly coherent pulsar wind that extends unperturbed up to the intra-binary shock \citep[see, e.g.][and references therein]{An2020, Sim2024}.
Unless the emission is strongly beamed towards us, even if scattered photons maintain the phase information on the pulsar spin and are emitted at once by the intra-binary shock, the size of this region cannot be too large.
Its displacement along the line of sight cannot exceed $c/\nu\approx 860$~km to make the light travel times of the photons from different locations in the reprocessing medium to the observer close enough to maintain coherence. The low duty cycle of the signal ($\simeq 1/32$) further reduces the maximum displacement to $\approx 30$~km. 

So far, observations of five isolated, young rotation-powered pulsars have revealed optical pulsations \citep[see, e.g.][and references therein]{Mignani2011}. These pulsars convert between $\eta \approx 10^{-8.5}$ and $10^{-5}$ of their spin-down power into pulsed emission observed in the B band (see black dots in Fig.~\ref{fig:fig3}). The optical emission of these pulsars is commonly ascribed to incoherent synchrotron radiation produced by relativistic particles accelerated in the neutron star magnetosphere \citep{Pacini1983}. Despite releasing a spin-down power more than four orders of magnitude weaker than the Crab pulsar and PSR J0540--6919, its twin in the Large Magellanic Cloud, the efficiency of {\psr} is close to theirs ($\eta=L_{\rm B}^{\rm pulse}/\dot{E}=3.3^{+2.4}_{-1.6}\times10^{-6}$; see the orange dot in Fig.~\ref{fig:fig3}) and suggests the same emission process. The observed efficiency is also of the same order as the upper limit set by \citet{strader2016} for PSR J0337+1715, an MSP in a hierarchical triple system with two white dwarfs. On the other hand, the transitional MSP PSR J1023+0038 \citep{Ambrosino2017} and the accreting MSP SAX J1808.4--3658 \citep{Ambrosino2021} are $\approx 50$--$100$ times more efficient, suggesting that an accretion disk around the neutron star may be crucial to enhance particle acceleration, as discussed in \citet{Papitto2019} and \citet{Veledina2019}.

The proposed optical pulsations from a redback rotation-powered MSP lend further support to the idea that fast rotating pulsars accelerate particles with high efficiency. In addition, the observations discussed here allow us to gauge the sensitivity requirement to search for signals from similar systems. The strength of the signal we reported is just above the sensitivity that can be reached during two consecutive nights of observations at a $\sim$4-m class telescope. Strikingly, the signal was observed when the source optical emission did not show the typical orbital modulation but rather flared close to the maximum magnitude reported for the source. 
Future quasi-simultaneous observations with radio and possibly even larger optical facilities will be key to increasing the significance of the detection and testing whether the pulsation strength is related to the overall variability properties of the optical light curve.

\begin{acknowledgements}
      This work was supported by INAF (Research Large Grant FANS, PI: Papitto), the Italian Ministry of University and Research (PRIN MUR 2020, Grant 2020BRP57Z, GEMS, PI: Astone), and 
Fondazione Cariplo/Cassa Depositi e Prestiti (Grant 2023-2560, PI: Papitto).
\end{acknowledgements}

\bibliographystyle{aa}
\bibliography{main.bib}

\begin{appendix}
\section{Noise distribution of SiFAP2 data}
\label{sec:app}
\begin{figure}[h!]
   \centering
   \includegraphics[width=\hsize]{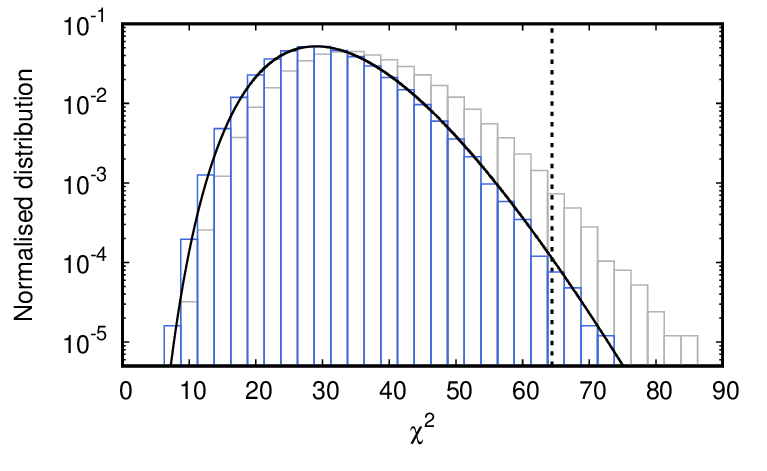}
      \caption{Grey boxes show the observed density of $\chi_{\rm tr}^2$ values observed epoch folding SiFAP2 W2 data around 100,000 trial periods close but not overlapping with the pulsar period. Blue boxes show the distribution of $\chi^2$ values rescaled by $r=1.16$, and the black solid line the $\chi^2_{n-1}$ probability density function for $n=32$. The vertical dashed line marks the rescaled value of $\chi^2_{\rm W2}=64.4$. }
         \label{fig:fig4}
   \end{figure}

\begin{figure}[h!]
   \centering
   \includegraphics[width=\hsize]{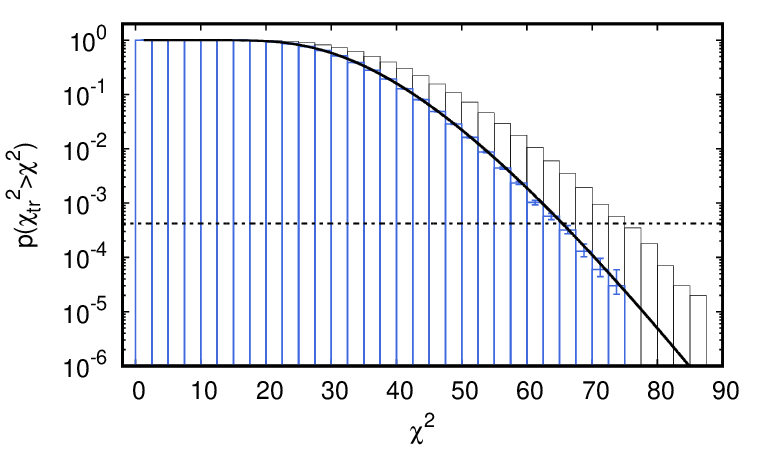}
      \caption{Grey boxes indicate the complementary cumulative distribution of the $\chi_{\rm tr}^2$ values obtained by epoch folding searching W2 data at 100,000 independent trial periods (see text for details). Blue boxes indicate the distribution obtained after rescaling the $\chi^2$ values by $r=1.16$. The expected $\chi^2$ tail distribution for $n=32$ is overplotted as a black line. The dashed horizontal line indicates the false alarm probability of the $\chi_{W2}^2=64.4$ value observed in SiFAP2 W2 data.     }
         \label{fig:fig5}
   \end{figure}

To study the distribution of noise powers and determine the false alarm probability of the signal observed in SiFAP2 W2 data, we performed 100,000 epoch folding trials using $n=32$ phase bins at periods evenly spaced at the independent Fourier resolution, and close to, but not overlapping with, the known period of J2339. Grey boxes in Fig.~\ref{fig:fig4} and Fig.~\ref{fig:fig5} show the normalised distribution and the cumulative probability of the observed $\chi_{\rm tr}^2$ values. Clearly, they do not follow the expected probability density and complementary cumulative distribution function for a $\chi^2_{n-1}$ distribution for $n=32$ (overplotted using black lines). The expected complementary cumulative distribution function for such a distribution, $1-P(n/2,x/2)$,  with $P(s,t)$ the regularised lower-incomplete gamma function, best fits the observed data when the $\chi_{tr}^2$ values are rescaled by a factor $r=1.16$ (see blue boxes in both figures). This factor is the total dispersion index of the distribution, i.e. the ratio between the observed variance of the distribution and its mean ($n-1$, in this case). \citet{LaPlaca2025} analysed a broader sample of observed SiFAP2 counting distributions to quantify the effect of crosstalk and found similar values. Rescaling the epoch folding $\chi^2$ obtained folding W2 data around the timing solution of the pulsar by this factor gave the values of $\chi^2$ quoted in text. For $n=32$, the rescaled value obtained in W2, $\chi^2_{\rm W2}=64.4$, has a probability of $p(\chi^2\geq\chi_{W2}^2)=4.0\times10^{-4}$ to be exceeded by noise. For comparison, the rescaled $\chi^2_{\rm tr}$ observed during the 100,000 trials at periods not overlapping with the pulsar period exceeded  $\chi_{W2}$ for a fraction of $(3.6^{+0.7}_{-0.5}\times10^{-4})$ of the total number of trials, in accordance with the value derived above.

\begin{figure}[h!]
   \centering
   \includegraphics[width=\hsize]{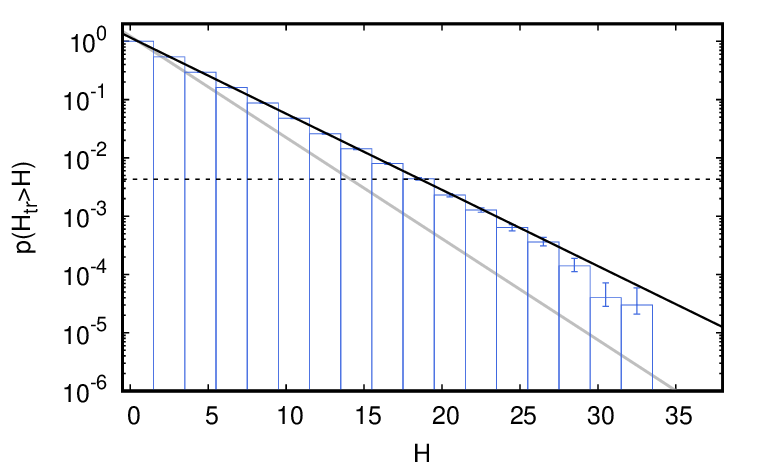}
      \caption{Complementary cumulative distribution of the $H_{\rm tr}^2$ values obtained by performing an H test on W2 data at 100,000 independent trial periods (see text for details). The black and grey solid lines indicate the best-fit function of the observed distribution, $p(H_{\rm tr}>H)=\exp{(-B\cdot H)}$ for $B=0.3$, and the relation expected if the noise is uncorrelated ($B=0.4$, \citet{deJager2010}), respectively. The horizontal dashed line marks the false alarm probability of the $H^*=18.1$ value observed in SiFAP2 W2 data.}
         \label{fig:fig6}
   \end{figure}

Similarly, to evaluate the false alarm probability of an H test in SiFAP2 W2 data, we evaluated the H statistics for the same  100,000 independent periods considered above. Fig.~\ref{fig:fig6} shows the complementary cumulative distribution of the observed $H_{\rm tr}$ values. The distribution is best fitted by a function $p(H_{\rm tr}>H)=\exp{(-B\cdot H)}$ with $B=-0.30$. This distribution is flatter than the expected distribution when the noise is uncorrelated \citep[B=0.4][]{deJager2010}, indicating how crosstalk slightly decreases SiFAP2 sensitivity. The false alarm probability of observing a value equal to, or higher than, $H^*=18.1$ (as for SiFAP2 W2 data; see text) is $4.3\times10^{-4}$. For comparison, the $H_{\rm tr}$ exceeded  $H^*$, $(4.2\pm0.2)\times10^{-3}$ times the total number of trials.

\end{appendix}
\end{document}